\documentclass[prb,amsfonts,showpacs,preprint]{revtex4}
\usepackage{hyperref,graphics,bm}
\begin{document}
\draft
\widetext
\title{Optimized Effective Potential for Extended Hubbard Model}
\author{I. V. Solovyev}
\address{
JRCAT-Angstrom Technology Partnership,
1-1-4 Higashi, Tsukuba, Ibaraki 305-0046, Japan\\
and
Institute of Metal Physics, Russian Academy of Sciences,
Ekaterinburg GSP-170, Russia
}
\date{\today}

\widetext
\begin{abstract}
Antiferromagnetic and charge ordered Hartree-Fock solutions of
the one-band Hubbard model with on-site and nearest-neighbor
Coulomb repulsions
are exactly mapped onto an auxiliary local Kohn-Sham (KS) problem
within a density-functional theory.
The mapping provides a new insight into
the interpretation of the KS equations.
(i) With an appropriate choice of the basic variable,
there is a universal form of the KS potential, which is applicable both
for
the antiferromagnetic and the charge ordered solutions.
(ii) The Kohn-Sham and Hartree-Fock eigenvalues are interconnected by
a scaling transformation.
(iii) The band-gap problem is attributed to the
derivative discontinuity of the basic variable as the function of the electron
number, rather than a finite discontinuity of the KS potential.
(iv) It is argued that the conductivity gap and the energies
of spin-wave excitations can be entirely defined by the parameters of the
ground state and the KS eigenvalues.
\end{abstract}

\pacs{71.15.Mb, 71.45.Gm, 71.45.Lr, 75.30.Fv}


\maketitle

\section{Introduction}

  Practical success of the Kohn-Sham (KS) scheme of density-functional theory (DFT)
\cite{KohnSham,ParrYang} in theoretical description of the ground-state
properties of molecules and solids lies in the idea of substitution of the real
interacting many-electron system with an auxiliary system of noninteracting
particles moving in an effective local potential. Similar ideas can be found
recently
in the dynamic mean-field theory of strongly correlated
systems, \cite{DMFT} which is based on a mapping of lattice models
(e.g., the Hubbard model or the Kondo model)
onto a local single-impurity
problem. This mapping is well elaborated only for a limited number of
examples like the homogeneous electron gas in the density-functional theory
[the so-called local-spin-density approximation (LSDA)], or in the limit
of infinite lattice coordination in the dynamic mean-field theory.
However, many general aspects
of such a local noninteracting representation
underlying the KS scheme are not well understood. One of the most
controversial issues is the physical meaning (if any) of
the KS eigenvalues, and their relationship with
true physical one-particle excitations.

  The major problem, which was intensively discussed in the 1980s,
\cite{Perdew1,Perdew2,ShamSchluter,GSS1,GSS2,HL,GunSch,SchGun,comment,review}
is the description of fundamental conductivity gaps in semiconductors and insulators.
It is generally accepted that the KS one-electron orbital energies have no
precise physical meaning, except for the highest occupied eigenvalue, which
is equal to the chemical potential (minus the work function).
Formally, the conductivity gap is not the ground-state property and includes
the excited-state effects. The LSDA systematically underestimate the band gaps.
The band-gap problem is typically ascribed to the finite discontinuity of
the KS potential as the function of the electron number, meaning that for an
extra electron, the KS potential may have an additional constant compared to
the KS potential designed for the ground state.
In other words, this is equivalent to the statement that the band-gap problem
for semiconductors and insulators may be resolved by a rigid upward shift of
the conduction bands in the KS picture for the ground state.
This procedure is frequently referred to as a ''scissor operator''.

  None of the standard density
functionals, such as the LSDA or generalized gradient expansions show such a
discontinuity. The origin and the magnitude of this discontinuity is still
an open question. Several authors argued that the discontinuity can be large,
and the single KS potential (even an exact one) designed for the ground state
does not give the correct gap.
\cite{Perdew2,ShamSchluter,GSS1,GSS2}
In accordance to their opinion,
the solution for the band-gap problem should be sought
in the nonlocality and the energy-dependence of the electron self-energy,
which is beyond the standard KS formulation, and can be
obtained, for example, in the framework of the $GW$ approximation.
\cite{GSS1,GSS2,HL,review}
On the contrary, Gunnarsson and Sch\"{o}nhammer suggested that the actual
discontinuity of the exact KS potential may be small, and the main problem lies
in the LSDA. \cite{GunSch} Their original arguments were based on a
numerical analysis of a Hubbard-type model with the on-site Coulomb
repulsion only. However, later they also argued that the intersite Coulomb
interaction can produce a large discontinuity of the KS potential. \cite{SchGun}
The disputed issues have been addressed again recently in Refs. \onlinecite{Arno}
and \onlinecite{SGN}, using two different viewpoints on the formulation of
the density functional-theory on a discrete lattice.

  In this work we discuss several general aspects of the density-functional
theory for the one-band extended Hubbard model.
We focus our attention on the analysis of
antiferromagnetic (AF)
and charge ordered insulating states, the description of which was
a traditionally difficult subject for {\it ab initio}
DFT methods. \cite{remark1,remark2,remark3}
From the very beginning we restrict ourselves by the framework of the (static)
mean-filed (MF) approximation.
Physically, this means that we
take into the consideration only the (classical) Coulomb and exchange
interactions among the electrons, and
neglect the correlation effects.
In this sense our density functional theory for the Hubbard model is
certainly less advanced rather than similar theories existing in the literature and
dealing with a formally exact problem numerically.
\cite{GunSch,SchGun,Arno,SGN} On the other hand, we hope that such a simplified
mean-filed analysis may be rather instructive, because
we will be able to treat the
problem analytically and give a clear picture for several disputed moments.
Our primary goal
will be to show how the same mean-field problem can be formulated in two
different forms, one of which is the conventional Hartree-Fock (HF) theory,
where the one-electron orbitals have a clear physical meaning based on
Koopmans' theorem, and the other
one is the KS scheme in the DFT.
The main point, which we would like to emphasize in this work is that the
meaning of the KS band structure may not be so formal, as it is currently
understood. The relationship between the KS eigenvalues and the true
one-particle excitations may be more fundamental, and is not necessarily based
only on the discontinuity property of the exact KS potential.
We do not discuss here the relative stability of different solutions of
the extended Hubbard model. Such an analysis can be found elsewhere
(see, e.g., Refs. \onlinecite{Hirsch,Brink,Meinders}
and references therein).

  The paper is organized as follows: the total energy functional is defined
in Sec.\ref{sec:totalenergy}. The KS potential is derived in Sec.\ref{sec:OEP}
by using the optimized effective potential (OEP) method. \cite{TalmanShadwick}
The connection between the OEP method, the density functional theory and
the unrestricted Hartree-Fock approach is discussed in Sections \ref{sec:Levy}
and \ref{sec:HF}. In Sec.\ref{sec:excitations} we discuss the connection
between
the conductivity gap, the energies of spin-wave excitations and the KS
one-electron energies.
Brief summary
will be given in Sec.\ref{sec:conclusion}.

\section{Total energy}
\label{sec:totalenergy}

  The one-band extended Hubbard Hamiltonian with on-site
($U$) and nearest-neighbor ($V$) Coulomb repulsions is defined as
\begin{equation}
\widehat{\cal H} = - t \sum_{\stackrel {\langle {\it ij} \rangle} {\sigma} }
\left( \widehat{c}^{\sigma \dagger}_{\it i} \widehat{c}^\sigma_{\it j} +
\widehat{c}^{\sigma \dagger}_{\it j} \widehat{c}^\sigma_{\it i} \right)
+ U \sum_{\it i} \widehat{n}_{\it i}^\uparrow  \widehat{n}_{\it i}^\downarrow +
V \sum_{ \stackrel{\langle {\it ij} \rangle} {\sigma, \sigma '} }
\widehat{n}_{\it i}^\sigma \widehat{n}_{\it j}^{\sigma '},
\label{eqn:EHM}
\end{equation}
where $\widehat{n}_{\it i}^\sigma$$=$$\widehat{c}^{\sigma \dagger}_{\it i} \widehat{c}^\sigma_{\it i}$
and $\widehat{c}^{\sigma \dagger}_{\it i}$ ($\widehat{c}^\sigma_{\it i}$) creates (annihilates)
an electron on the site ${\it i}$ with spin $\sigma$$=$$\uparrow$ or $\downarrow$.
A nearest-neighbor pair is denoted by $\langle {\it ij} \rangle$. The kinetic hopping between
nearest-neighbor orbitals is denoted by $t$.
We use the convention, in accordance with which
capital letters
are reserved for absolute values of parameters of Coulomb interactions and
the KS potential, whereas small letters
stand for the same parameters in units of $2t$
(for example, $u$$=$$U/2t$, etc.).

  Let us consider the mean-field solution of the extended Hubbard model,
where the ground-state wavefunction of the many-electron system is
approximated by a single Slater determinant $|S[\psi^\sigma_m]\rangle$ of
one-electron orthonormal orbitals $\{\psi^\sigma_m\}$.
Then, the total energy is given by
\begin{equation}
{\cal E}_{\rm MF}[\psi^\sigma_m] = \langle S[\psi^\sigma_m]|
\widehat{\cal H} | S[\psi^\sigma_m]\rangle,
\label{eqn:te1}
\end{equation}
and can be expressed in terms of the one-electron density matrix
\begin{equation}
\rho^\sigma[{\it i},{\it j}]=\sum_m \langle \psi^\sigma_m |
\widehat{c}^{\sigma \dagger}_{\it i} \widehat{c}^\sigma_{\it j} |
\psi^\sigma_m \rangle
\label{eqn:dmatrix}
\end{equation}
as ${\cal E}_{\rm MF}$$=$${\cal T}$$+$${\cal E}_{\rm H}$$
+$${\cal E}_{\rm X}$ (see, e.g., Ref. \onlinecite{ParrYang}),
where ${\cal T}$ is the (noninteracting) kinetic
energy:
\begin{equation}
{\cal T}=-t \sum_{\stackrel {\langle {\it ij} \rangle} {\sigma} }
\left(\rho^\sigma[{\it i},{\it j}]+\rho^\sigma[{\it j},{\it i}]\right),
\label{eqn:ekin}
\end{equation}
${\cal E}_{\rm H}$ is the Hartree (Coulomb) energy:
\begin{equation}
{\cal E}_{\rm H}=\frac{U}{2}\sum_{\it i} \left(\rho^\uparrow[{\it i},{\it i}]
+ \rho^\downarrow[{\it i},{\it i}]\right)^2 + V
\sum_{\langle {\it ij} \rangle} \left(
\rho^\uparrow[{\it i},{\it i}] + \rho^\downarrow[{\it i},{\it i}]\right)
\left(
\rho^\uparrow[{\it j},{\it j}] + \rho^\downarrow[{\it j},{\it j}]\right),
\label{eqn:ehartree}
\end{equation}
and ${\cal E}_{\rm X}$ is the exchange (Fock) energy:
\begin{equation}
{\cal E}_{\rm X} = -\frac{U}{2}\sum_{{\it i} \sigma}
\left(\rho^\sigma[{\it i},{\it i}]\right)^2 - V
\sum_{\stackrel {\langle {\it ij} \rangle} {\sigma} }
\rho^\sigma[{\it i},{\it j}]\rho^\sigma[{\it j},{\it i}].
\label{eqn:efock}
\end{equation}
Both ${\cal E}_{\rm H}$ and ${\cal E}_{\rm X}$ include the self-interaction.

\begin{figure}
\centering \noindent
\resizebox{12cm}{!}{\includegraphics{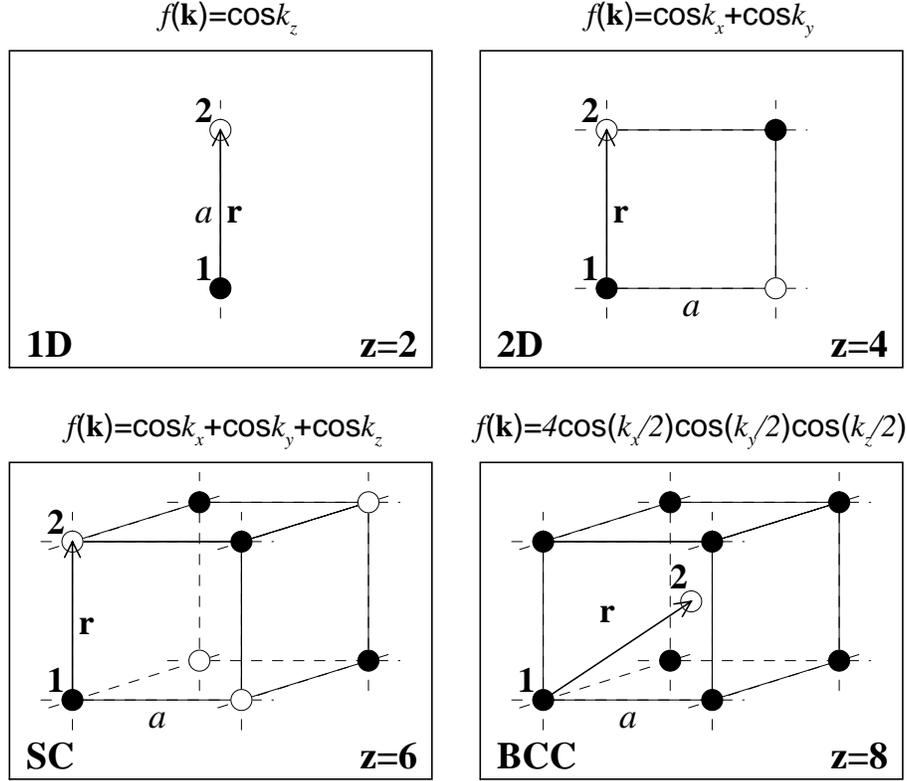}}
  \caption{Doubling of the unit cell associated with the
           spin or/and charge ordering. Two nonequivalent sites
           ($\nu$$=$$1,2$) are shown by white and black circles.
           Within a doubled unit cell, the site 1 is located in
           the origin and the position of the site 2 is defined by
           the vector ${\bf r}$. $z$ is the coordination number.
           $a$ is the lattice parameter.
           $f({\bf k})$ is the structure factor - see text
           (${\bf k}$ is in the units of $1/a$).}
  \label{fig.lattices}
\end{figure}

  Our goal is to simulate a spin or/and charge ordering on the bipartite
one-dimensional (1D), two-dimensional (2D), simple cubic (SC) or
base-centered cubic (BCC) lattice, for which the spin or/and charge
densities are periodically distributed between two types of atoms
$\nu$$=$ 1, 2
resulting in the doubling of the unit cell, as it is shown in Fig.\ref{fig.lattices}.
Thus, the site-indices are specified by the position of the
doubled unit cell ${\bf R}$ and the type of the site:
${\it i}$$\equiv$$({\bf R} \nu)$.
$\rho^\sigma$ is the periodic object:
$\rho^\sigma[{\bf R}'\nu',{\bf R}\nu]$$=$$\rho^\sigma[({\bf R}'-{\bf R})\nu',0\nu]$,
implying that the one-electron orbitals in Eq.(\ref{eqn:dmatrix}) are the Bloch
states: i.e., $m$$\equiv$$({\bf k}\ell)$ and
$\widehat{c}^\sigma_{{\bf R} \nu}|\psi^\sigma_{{\bf k}\ell}\rangle$$=$$
e^{i{\bf kR}}\widehat{c}^\sigma_{0 \nu}|\psi^\sigma_{{\bf k}\ell}\rangle$,
where
${\bf k}$ is a vector in the first Brillouin zone of the doubled
unit cell, and $\ell$ is the remaining (band) index.
In accordance with the definition (\ref{eqn:EHM}), the range of interactions
in the Hamiltonian
$\widehat{\cal H}$ is restricted by the nearest neighbors.
If we combine this fact with the symmetry properties of
the density matrix, i.e., the periodicity and the invariance with respect
to the local symmetry operations for the lattices shown in Fig.\ref{fig.lattices},
it is rather easy to see that the total energy (\ref{eqn:te1}) will be
entirely represented by the elements of the density matrix with
${\bf R}$$=$${\bf R}'$$=0$, which can be formally composed into the
$2$$\times$$2$ matrix:
\begin{equation}
\widehat{\rho}^\sigma = \left( \begin{array}{cc}
\rho^\sigma [01,01] & \rho^\sigma [01,02] \\
\rho^\sigma [02,01] & \rho^\sigma [02,02] \\
\end{array} \right).
\label{eqn:dmatrix1}
\end{equation}
Since the inversion is one of the local symmetry operations,
$\widehat{\rho}^\sigma$
may be taken in the real
symmetric form.

  In the following we
will focus out attention on the three possible
ordered states: \\
the antiferromagnetic state near to half filling,\\
the charge-ordered nonmagnetic (CON) state near to half filling,\\
the charge-ordered ferromagnetic (COF) state near to quarter filling.\\
The number of electrons in the doubled unit cell is specified as
$n_e$$=$$n_0$$\pm$$\eta$, where $n_0$ is the integer electron number at
half/quarter filling ($n_0$$=$$2$ for the AF and CON state, and $n_0$$=$$1$
for the COF state) and $\eta$ describes a small deviation away from
half/quarter filling.
Then, the elements of the density matrix
for these three cases
obey the series of symmetry constraints listed in Table \ref{constraints},
reducing the
number of independent elements in
$\widehat{\rho}^{\uparrow, \downarrow}$, for a given $n_e$,
to the expectation values of only
two pseudospins calculated for the majority-spin channel:
\begin{equation}
\tau_{\rm x} = {\rm Tr}(\widehat{\rho}^\uparrow\widehat{\sigma}_{\rm x}), \ \ \
\tau_{\rm z} = {\rm Tr}(\widehat{\rho}^\uparrow\widehat{\sigma}_{\rm z}),
\label{eqn:pseudospins}
\end{equation}
where $\widehat{\sigma}_x$ and $\widehat{\sigma}_z$ are $2$$\times$$2$ Pauli
matrices.
In the AF state, $\tau_{\rm z}$
is equal to the local magnetic moment.
If we define the number of the doubled unit cells as $N$, and the total energy
per doubled unit cell as $E_{\rm MF}$$\equiv$${\cal E}/N$$=$$T$$+$$E_{\rm H}$$+$$E_{\rm X}$,
we obtain from Eqs.(\ref{eqn:ekin})-(\ref{eqn:efock}) the following expressions for the
kinetic energy ($T$), the Hartree energy ($E_{\rm H}$), and the exchange energy
($E_{\rm X}$) per doubled unit cell, correspondingly
\begin{equation}
T [\tau_{\rm x}] = - n_0 z t \tau_{\rm x},
\label{eqn:Ekin}
\end{equation}
\begin{equation}
E_{\rm H} [\tau_{\rm z}] = \frac{1}{4} n_e^2 (U+zV) + \frac{1}{4} (U-zV)
(\tau_{\rm z} + \tau_{\rm z}^\downarrow)^2,
\label{eqn:EH}
\end{equation}
and
\begin{equation}
E_{\rm X}[\tau_{\rm x},\tau_{\rm z}] = -\frac{1}{4} n_0 U
\left[ \left( \frac{n_e}{n_0}\right)^2 + \tau_{\rm z}^2 \right]
 -\frac{1}{4} n_0 zV \tau_{\rm x}^2,
\label{eqn:EX}
\end{equation}
where $z$ is the coordination number. In Eq.(\ref{eqn:EH}),
$\tau_{\rm z}^\downarrow$ is $-$$\tau_{\rm z}$, $\tau_{\rm z}$ and
$0$ for the AF, CON and COF states, respectively.
The prefactor $n_0$ in Eqs.(\ref{eqn:Ekin}) and (\ref{eqn:EX}) counts
the spin contributions. Due to the symmetry properties of the density matrix
(Table \ref{constraints}), we have to equivalent contributions to $T$ and
$E_{\rm X}$ coming from the majority-spin and minority-spin channels in the
case of AF and CON ordering. This is reflected in the prefactor $n_0$$=$$2$,
which coincides with the number of electrons at half filling. In the case of
COF ordering, the minoriti-spin states are unoccupied. Therefore, we have to
count only the majority-spin channel and take $n_0$$=$$1$, which again
concides with the number of electrons at quarter filling.

\section{Optimized effective potential}
\label{sec:OEP}

  There are two possible ways of searching the ground state for
the mean-field total energy functional (\ref{eqn:te1}).
One is the standard Hartree-Fock approach, which will be discussed in Sec.\ref{sec:HF}.
Here we start with an alternative provided by the KS scheme and
assume that the one-electron orbitals
$\{ \psi^\sigma_{{\bf k} \ell} \}$ can be obtained as $n_e$$\times$$N$ lowest energy
eigenfunctions of the following Hamiltonians:
\begin{equation}
\widehat{\cal H}_{\rm KS}^\sigma = -t \sum_{\langle{\bf R}1{\bf R}'2\rangle}
\left( \widehat{c}^{\sigma \dagger}_{{\bf R} 1}\widehat{c}^\sigma_{{\bf R}'2} +
\widehat{c}^{\sigma \dagger}_{{\bf R}'2}\widehat{c}^\sigma_{{\bf R} 1} \right) +
\sum_{\bf R} \left( B^\sigma_1 \widehat{c}^{\sigma \dagger}_{{\bf R} 1}\widehat{c}^\sigma_{{\bf R} 1}
+ B^\sigma_2 \widehat{c}^{\sigma \dagger}_{{\bf R} 2}\widehat{c}^\sigma_{{\bf R} 2} \right).
\label{eqn:HKS}
\end{equation}
As in Eq.(\ref{eqn:EHM}), the notation $\langle{\bf R}1{\bf R}'2\rangle$ means that the
summation is restricted by the nearest neighbors, which are now of different types
(see Fig.\ref{fig.lattices}). If ${\bf r}$ is the relative position of two different
atoms within one unit cell, the notation $\langle{\bf R}1{\bf R}'2\rangle$ is
equivalent to the condition $|{\bf R}'+{\bf r}-{\bf R}|$$=$$|{\bf r}|$.
Thus, $\widehat{\cal H}_{\rm KS}^{\uparrow,\downarrow}$ retains the kinetic part of
by the original Hamiltonian $\widehat{\cal H}$.
The interacting part is substituted
by the effective one-electron KS potentials $\{ B_\nu^\sigma \}$, which
are assumed to be site-diagonal (''local'' with respect to the site-indices),
do not depend on ${\bf R}$ (as it is requested by the periodicity of the
antiferromagnetic and charge-ordered states), but may depend on the type of the
atom $\nu$.  $\{ B_\nu^\sigma \}$
can be found in the spirit of optimized effective potential method
\cite{TalmanShadwick,Gross}
as the solution of the variational problem (for the given electron number
$n_e$):
\begin{equation}
\frac {\partial}{\partial B_\nu^\sigma} \left. \left(
T[\tau_{\rm x}]+E_{\rm H}[\tau_{\rm z}]+E_{\rm X}[\tau_{\rm x},\tau_{\rm z}]
 - \sum_\lambda \lambda {\cal F}_\lambda
[ \{ B_\nu^\sigma \} ] \right) \right|_{n_e} = 0,
\label{eqn:OEP}
\end{equation}
provided that $\{ B_\nu^\sigma \}$ obey
the series of constraints ${\cal F}_\lambda[ \{ B_\nu^\sigma \} ]$,
which guarantee the
required symmetry of the density matrix (Table \ref{constraints}).
Thus, $\{ B_\nu^\sigma \}$ can be interpreted as the exact KS potentials
for exchange alone. \cite{ParrYang}
The Hamiltonian (\ref{eqn:HKS}) can be diagonalized in the basis of
Bloch orbitals
\begin{equation}
\left( \begin{array}{r}
\widehat{c}^{\sigma \dagger}_1 ({\bf k}) \\
\widehat{c}^{\sigma \dagger}_2 ({\bf k}) \\
\end{array} \right)
 = \frac{1}{\sqrt{N}} \sum_{\bf R} e^{i\bf{kR}}
\left( \begin{array}{r}
\widehat{c}^{\sigma \dagger}_{{\bf R}1} \\
e^{i\bf{kr}}\widehat{c}^{\sigma \dagger}_{{\bf R}2} \\
\end{array} \right).
\label{eqn:basis}
\end{equation}
The eigenfunctions of $\widehat{\cal H}_{\rm KS}^\sigma$ are
\begin{equation}
\left( \begin{array}{c}
| \psi^\sigma_{{\bf k}-} \rangle \\
| \psi^\sigma_{{\bf k}+} \rangle \\
\end{array} \right)
=
\left( \begin{array}{rr}
 \cos \vartheta^\sigma_{\bf k}  &  \sin \vartheta^\sigma_{\bf k} \\
-\sin \vartheta^\sigma_{\bf k}  & \cos \vartheta^\sigma_{\bf k}  \\
\end{array} \right)
\left( \begin{array}{c}
\widehat{c}^{\sigma \dagger}_1 ({\bf k}) | \rangle \\
\widehat{c}^{\sigma \dagger}_2 ({\bf k}) | \rangle \\
\end{array} \right),
\label{eqn:eigenf}
\end{equation}
where $| \rangle$ denotes the vacuum state with no electrons in it.

  We note that due to the symmetry constraints (Table \ref{constraints}),
we have to consider only the case
$-$$B_1^\uparrow$$=$$B_2^\uparrow$$\equiv$$B$. Then, the eigenvalues of
$\widehat{H}_{\rm KS}^\uparrow$ are
\begin{equation}
\varepsilon^\uparrow_\pm ({\bf k})=\pm
\sqrt{B^2+4t^2 f^2({\bf k})}
\label{eqn:eigenv}
\end{equation}
and the angle $\vartheta^\uparrow_{\bf k}$ in Eq.(\ref{eqn:eigenf}) is
given by
\begin{equation}
{\rm tg} \vartheta^\uparrow_{\bf k}=\frac{f({\bf k})}
{b+\sqrt{b^2+f^2({\bf k})}},
\label{eqn:angle}
\end{equation}
where
$$f({\bf k})=\frac{1}{2}\sum_{{\bf R}:|{\bf r}-{\bf R}|=|{\bf r}|} \cos {\bf k}
({\bf r}-{\bf R})$$
is the structure factor.
[The explicit from of $f({\bf k})$ for
different lattices is shown in Fig.\ref{fig.lattices}].
Then,
putting the wavefunctions (\ref{eqn:eigenf}) in equations (\ref{eqn:dmatrix}),
(\ref{eqn:dmatrix1}) and (\ref{eqn:pseudospins}),
we obtain after some algebra:
\begin{equation}
\tau_{\rm x} = \frac {2}{zN} \sum_{{\bf k}\ell}
\frac {(-\ell)f^2({\bf k})}
{\sqrt{b^2 + f^2({\bf k})}}
\Theta[\varepsilon_F-\varepsilon^\uparrow_\ell({\bf k})],
\label{eqn:taux}
\end{equation}
\begin{equation}
\tau_{\rm z} = \frac {1}{N} \sum_{{\bf k}\ell}
\frac {(-\ell)b}
{\sqrt{b^2 + f^2({\bf k})}}
\Theta[\varepsilon_F-\varepsilon^\uparrow_\ell({\bf k})],
\label{eqn:tauz}
\end{equation}
where $\ell$$=$$\pm$$1$, and the Fermi energy $\varepsilon_F$ is determined by
the condition:
\begin{equation}
n_0 \sum_{{\bf k}\ell} \Theta[\varepsilon_F-\varepsilon^\uparrow_\ell({\bf k})]
=n_eN.
\label{eqn:efermi}
\end{equation}
Then, it is rather straightforward to show for any fixed electron number
$n_e$ that
\begin{equation}
\left. \frac {\partial \tau_{\rm x}}{\partial b} \right|_{n_e} =
- \frac {2b}{z} \left. \frac {\partial \tau_{\rm z}}{\partial b} \right|_{n_e}.
\label{eqn:xzcoupling}
\end{equation}
Therefore, $\tau_{\rm x}$ and $\tau_{\rm z}$ are
not independent variables and the variational procedure of the OEP approach
is equivalent to a DFT
in which $\tau_{\rm z}$ participates
as the basic variable. Indeed, using Eqs.(\ref{eqn:Ekin}) and (\ref{eqn:xzcoupling})
we obtain
$(\partial T[\tau_{\rm x}]/\partial\tau_{\rm z})|_{n_e}$$=$$n_0B$.
As it was discussed earlier, the prefactor $n_0$ counts the spin channels
involved into the problem.
Then, variational principle (\ref{eqn:OEP}) leads to the standard definition for the KS
potential conjugated with the basic variable $\tau_{\rm z}$:
\begin{equation}
B=-\frac{1}{n_0} \left\{ \left.
\frac{\partial E_{\rm H}[\tau_{\rm z}]}{\partial\tau_{\rm z}}\right|_{n_e}+
\left.
\frac{\partial E_{\rm X}[\tau_{\rm x},\tau_{\rm z}]}
{\partial \tau_{\rm z}}\right|_{n_e}\right\},
\label{eqn:OEPKS}
\end{equation}
where the
minus sign
is related with our definition of $\tau_{\rm z}$ and $B$:
for $\sigma$$=$$\uparrow$, the solution with
$\tau_{\rm z}$$=$$\rho^\uparrow[01,01]$$-$$\rho^\uparrow[02,02]$$>$$0$
corresponds to the potential $B^\uparrow_1$$=$$-$$B^\uparrow_2$$\equiv$$-$$B$$<$$0$.
The exchange part in Eq.(\ref{eqn:OEPKS}) is given by
\begin{equation}
B_{\rm X} \equiv
- \frac{1}{n_0}
\left. \frac{\partial E_{\rm X}[\tau_{\rm x},\tau_{\rm z}]}{\partial \tau_{\rm z}}\right|_{n_e}
=\frac{1}{2}
U \tau_{\rm z}-V \tau_{\rm x} b
\label{eqn:bx1}
\end{equation}
[see Eqs.(\ref{eqn:EX}) and (\ref{eqn:xzcoupling})].
The Coulomb
part $B_{\rm H}$$\equiv$$
(-1/n_0)(\partial E_{\rm H}[\tau_{\rm z}]/\partial \tau_{\rm z})|_{n_e}$
is $0$, $(U$$-$$zV) \tau_{\rm z}$ and
$\frac{1}{2} (U$$-$$zV) \tau_{\rm z}$ for the AF, CON and COF states,
respectively [see Eq.(\ref{eqn:EH})]. Then, Eq.(\ref{eqn:OEPKS}) can be transformed to
the self-consistent
equation for the
KS potential $b$$=$$b_{\rm H}$$+$$b_{\rm X}$:
\begin{equation}
b = \frac{1}{2} u_{\rm HF} \tau_{\rm z} - v \tau_{\rm x} b,
\label{eqn:SCF1}
\end{equation}
where $u_{\rm HF}$ is
$u$, $2zv-u$ and $zv$ for the AF, CON and COF states, respectively.
Eq.(\ref{eqn:SCF1}) further yields:
\begin{equation}
b = \frac{1}{2} u_{\rm KS}[\tau_{\rm x}] \tau_{\rm z},
\label{eqn:SCF2}
\end{equation}
where the effective coupling $u_{\rm KS}[\tau_{\rm x}]$ is given by
\begin{equation}
u_{\rm KS}[\tau_{\rm x}] = \frac{u_{\rm HF}}{1+v\tau_{\rm x}}.
\label{eqn:uks}
\end{equation}
Putting Eq.(\ref{eqn:SCF2}) in Eq.(\ref{eqn:bx1}), one can find for
the self-consistent
exchange potential $b_{\rm X}$:
\begin{equation}
b_{\rm X}=\frac{1}{2} \left[ \frac{v \tau_{\rm x}}{1+v \tau_{\rm x}}u_{\rm HF}
-u \right] \tau_{\rm z}.
\label{eqn:bx2}
\end{equation}

  The local KS potential has the same form [Eq.(\ref{eqn:SCF2})]
for different ordered states, although the driving force for the ordering
may be different: $u$ in the AF state and $v$ in the
CO states.
The local form of the KS potential for the CO states has one important
implication which we would like to emphasize here. It is well established
that Hubbard-type models with solely on-site Coulomb interactions treated
on the level of mean-field theories may exhibit the charge-ordered solutions
(see, e.g., Ref.\onlinecite{COmodels}). The present work suggests that such
''on-site Coulomb interactions'' may be interpreted in the Kohn-Sham sense,
as parameters of an auxiliary local KS problem, whose real physical meaning
may be different and implies the contributions of longer-range Coulomb
interactions.

\begin{figure}
\centering \noindent
\resizebox{12cm}{!}{\includegraphics{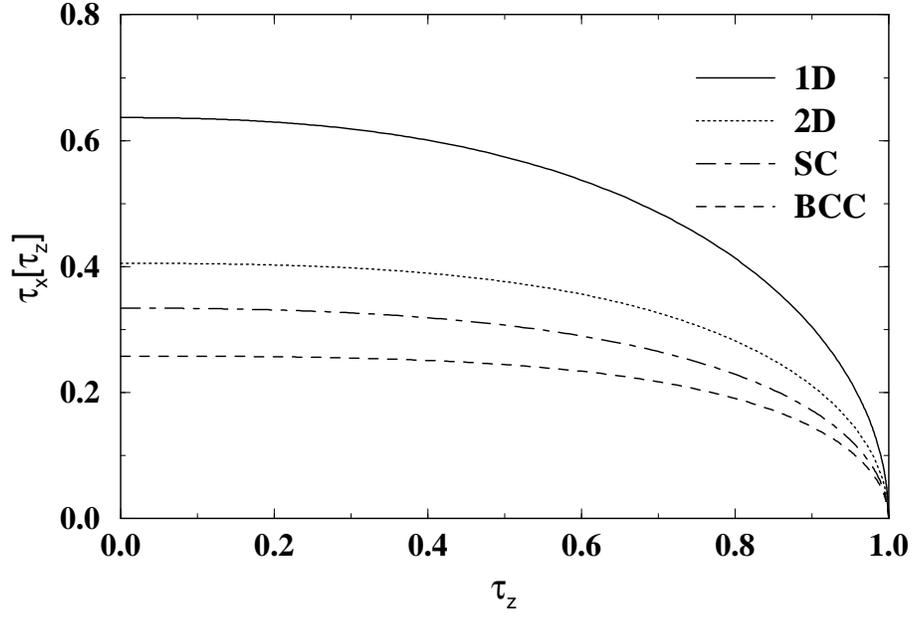}}
  \caption{$\tau_{\rm x}$ versus $\tau_{\rm z}$ for various lattices
           and $n_e$$=$$n_0$.}
  \label{fig.taux}
\end{figure}

  $u_{\rm KS}$ has the common renormalization factor
$(1$$+$$v\tau_{\rm x})^{-1}$ which is driven by the intersite
Coulomb interaction $v$ and depends on
$\tau_{\rm z}$ through $\tau_{\rm x}$.
The dependence $\tau_{\rm x}[\tau_{\rm z}]$ is shown in
Fig.\ref{fig.taux} for several lattices and $n_e$$=$$n_0$.
We note: (i) for a given $\tau_{\rm z}$, $\tau_{\rm x}$ is larger
when $z$ is smaller.
Thus, the renormalization
is the most important for 1D systems. (ii)
In the framework of our model, $\tau_{\rm z}$$=$$0$ may be regarded as a
homogeneous solution. \cite{remark1}
Since $\tau_{\rm x}$ is the
decreasing function of $\tau_{\rm z}$,
we have $u_{\rm KS}[0]$$\leq$$u_{\rm KS}[\tau_{\rm z}]$ meaning
that the parameter $u_{\rm KS}$ being once evaluated in the homogeneous limit
and employed afterwards for the whole range of $\tau_{\rm z}$ is generally
''overscreened''. It is therefore clear that an attempt to substitute
the real effective coupling $u_{\rm KS}[\tau_{\rm z}]$ with its
homogeneous counterpart $u_{\rm KS}[0]$
in the KS potential
will generally
underestimate the tendencies towards the antiferromagnetism and the
charge ordering, as it is well established for the LSDA. \cite{remark2}
The analytical estimate for $\tau_{\rm z}$$=$$0$ and
$n_e$$=$$n_0$ is
$\tau_{\rm x}[0]$$=$$ (2/\pi)^{\log_2z}$ (except the SC lattice).
At the opposite end $\tau_{\rm z}$$=$$1$, corresponding to the limit
$u_{\rm KS}[1]$$\rightarrow$$\infty$, we have
$\tau_{\rm x}$$\rightarrow$$1/2b$$
\rightarrow$$1/u_{\rm KS}[1]$, see Eqs.(\ref{eqn:taux}) and (\ref{eqn:SCF2}).
Then,
from Eq.(\ref{eqn:uks}) we obtain
$u_{\rm KS}[1]$$=$$u_{\rm HF}(1$$+$$v/u_{\rm KS}[1])^{-1}$, whence it
follows that
\begin{equation}
u_{\rm KS}[1]=u_{\rm HF}-v.
\label{eqn:inflimit}
\end{equation}

\section{The Levy constrained-search formulation}
\label{sec:Levy}

  A very elegant formulation of the density-functional theory was
introduced by Levy. \cite{ParrYang} For our purposes it is read as
follows: if $\tau_{\rm z}$ is the basic variable, the
total energy minimization (for a given electron number)
can be performed in two steps:
\begin{equation}
{\cal E} = \min_{\tau_{\rm z}} \left\{ \min_{|\Psi \rangle \rightarrow
\tau_{\rm z}} \langle \Psi | \widehat{\cal H} | \Psi \rangle \right\}.
\label{eqn:Levy1}
\end{equation}
The inner minimization is constrained to all many-electron
wavefunctions $| \Psi \rangle$ that give $\tau_{\rm z}$, while the
outer minimization releases the constraint by searching all
$\tau_{\rm z}$. In the mean-filed approximation,
the search of all possible
many-electron wavefunctions is restricted within the subspace
of Slater determinants $|S[\psi^\sigma_m]\rangle$.
Thus, $| \Psi \rangle$ is additionally constrained, and
we have ${\cal E}$$\le$${\cal E}_{\rm MF}$.
The purpose of this section is to show that in all other respects, the OEP
approach employed in the previous section is equivalent to the
Levy constrained-search formulation (\ref{eqn:Levy1}). The proof is
simple and can be carried out along the general line taken in the
standard DFT formulation. \cite{ParrYang}
Taking into account the explicit form of
$\langle S[\psi^\sigma_m] | \widehat{\cal H} | S[\psi^\sigma_m] \rangle$,
Eqs.(\ref{eqn:Ekin})-(\ref{eqn:EX}), the inner step in
Eq.(\ref{eqn:Levy1})
is equivalent to the search of the one-electron
orbitals $\{\psi^\sigma_m\}$ which
maximize the expectation value of the pseudospin $\tau_{\rm x}$
for the given $\tau_{\rm z}$, or equivalently,
minimize the (noninteracting)
kinetic energy $T$ for the given $\tau_{\rm z}$. Using the
definitions of $T$ and $\tau_{\rm z}$ in terms of the one-particle orbitals
[see Eqs. (\ref{eqn:dmatrix}), (\ref{eqn:ekin}), (\ref{eqn:dmatrix1}) and
(\ref{eqn:pseudospins})], this gives the functional
\begin{equation}
\Omega[\psi^\uparrow_m] = -t
\sum_m \sum_{\langle {\bf R} 1 {\bf R}'2 \rangle}
\langle \psi^\uparrow_m |
\widehat{c}^{\uparrow \dagger}_{{\bf R} 1}\widehat{c}^\uparrow_{{\bf R}'2} +
\widehat{c}^{\uparrow \dagger}_{{\bf R}'2}\widehat{c}^\uparrow_{{\bf R} 1}
| \psi^\uparrow_m \rangle -
\lambda^\uparrow \sum_m \sum_{\bf R} \left(
\langle \psi^\uparrow_m |
\widehat{c}^{\uparrow \dagger}_{{\bf R} 1}\widehat{c}^\uparrow_{{\bf R} 1} -
\widehat{c}^{\uparrow \dagger}_{{\bf R} 2}\widehat{c}^\uparrow_{{\bf R} 2}
| \psi^\uparrow_m \rangle - \tau_{\rm z} \right) -
\sum_{mm'} \varepsilon^\uparrow_{mm'}
\left( \langle \psi^\uparrow_m | \psi^\uparrow_{m'} \rangle
 - \delta_{mm'} \right),
\label{eqn:Levy3}
\end{equation}
and the minimization condition
$\delta \Omega[\psi^\uparrow_m] / \delta \langle \psi^\uparrow_m |$$=$$0$.
(The Lagrange multipliers $\varepsilon^\uparrow_{mm'}$ stand for the
orthonormalization condition between $| \psi^\uparrow_m \rangle$
and $| \psi^\uparrow_{m'} \rangle$). These result in the equation
\begin{equation}
\widehat{h}^\uparrow | \psi^\uparrow_m \rangle =
\sum_{m'} \varepsilon^\uparrow_{mm'} | \psi^\uparrow_{m'} \rangle,
\label{eqn:Levy4}
\end{equation}
where
\begin{equation}
\widehat{h}^\uparrow = -t
\sum_{\langle {\bf R} 1 {\bf R}'2 \rangle}
\left( \widehat{c}^{\uparrow \dagger}_{{\bf R} 1}\widehat{c}^\uparrow_{{\bf R}'2} +
\widehat{c}^{\uparrow \dagger}_{{\bf R}'2}\widehat{c}^\uparrow_{{\bf R} 1}
\right) - \lambda^\uparrow
\sum_{\bf R} \left(\widehat{c}^{\uparrow \dagger}_{{\bf R} 1}\widehat{c}^\uparrow_{{\bf R} 1} -
\widehat{c}^{\uparrow \dagger}_{{\bf R} 2}\widehat{c}^\uparrow_{{\bf R} 2}
\right).
\label{eqn:Levy5}
\end{equation}
Similar equations can be obtained for $\sigma$$=$$\downarrow$ by using the
symmetry properties of the density matrix from Table \ref{constraints}.
The Hamiltonian (\ref{eqn:Levy5}) has the same form as the KS Hamiltonian
(\ref{eqn:HKS}). (Note also that since $\widehat{h}^\uparrow$ is Hermitian,
$\varepsilon^\uparrow_{mm'}$ can be taken in the diagonal form).
If $\tau_{\rm z}$ is the expectation value of the pseudospin corresponding
to the ground state, the Lagrange multiplier $\lambda^\uparrow$ is equal to the
self-consistent potential $B$.
Thus, the inner minimization of the Levy constrained-search formulation is
equivalent to the diagonalization of the KS Hamiltonian in the OEP approach.

  The conclusion is very natural and, in principle, directly follows from the definition
of the KS scheme. \cite{KohnSham,ParrYang} Indeed, the mean-field energy
${\cal E}_{\rm MF}$ is determined by only two parameters,
$\tau_{\rm x}$ and $\tau_{\rm z}$. Both of them are requested to
come out exactly of the one-electron KS orbitals $\{\psi^\sigma_m\}$:
$\tau_{\rm z}$ as the basic variable, and $\tau_{\rm x}$ as the only
variable which determines the noninteracting kinetic energy.

\section{Unrestricted Hartree-Fock approach}
\label{sec:HF}

  In order to clarify the origin of the renormalization factor
$(1$$+$$v\tau_{\rm x})^{-1}$ in the KS potential (\ref{eqn:SCF2}),
let us now turn to the Hartree-Fock solution of the
extended Hubbard model, where the basic variables are
the one-electron orbitals
$\{\psi^\sigma_m\}$:
$\widehat{\cal H}^\sigma_{\rm HF} |\psi^\sigma_m \rangle$$=$$
\delta {\cal E}_{\rm MF}/\delta \langle \psi^\sigma_m|$.
Taking into account the periodicity of the ordered states,
this gives the Hamiltonian (the uniform potential shift is dropped):
\begin{equation}
\widehat{\cal H}^\uparrow_{\rm HF} = -t(1+v\tau_{\rm x})
\sum_{\langle {\bf R} 1 {\bf R}'2 \rangle}
\left( \widehat{c}^{\uparrow \dagger}_{{\bf R} 1}\widehat{c}^\uparrow_{{\bf R}'2} +
\widehat{c}^{\uparrow \dagger}_{{\bf R}'2}\widehat{c}^\uparrow_{{\bf R} 1}
\right) - \frac{1}{2}U_{\rm HF} \tau_{\rm z}
\sum_{\bf R} \left(\widehat{c}^{\uparrow \dagger}_{{\bf R} 1}\widehat{c}^\uparrow_{{\bf R} 1} -
\widehat{c}^{\uparrow \dagger}_{{\bf R} 2}\widehat{c}^\uparrow_{{\bf R} 2}
\right)
\label{eqn:HHF}
\end{equation}
with the parameters of site-diagonal interactions
$U_{\rm HF}$ defined earlier.
The prefactor $(1$$+$$v\tau_{\rm x})$ in the off-diagonal part of the HF Hamiltonian
comes from the nonlocality of the Fock exchange
potential, which is combined with the kinetic term. Since the KS potential
is requested to be local, the same effect enters the
KS formulation through the renormalization of the
site-diagonal interactions, as it is indeed manifested in Eq.(\ref{eqn:uks}).
Therefore, the terminology ''the local potential''
and ''the local approximation'' used in the density-functional theory
does not necessarily mean that the nonlocal exchange interactions
are totally neglected.
They may contribute to the renormalization of the local interactions.

  In the framework of the one-band extended Hubbard model,
we obtain the trivial connection between the KS and HF Hamiltonians:
$\widehat{\cal H}_{\rm KS}^\uparrow$$=$$(1$$+$$v\tau_{\rm x})^{-1}
\widehat{\cal H}_{\rm HF}^\uparrow$, meaning that they have
the same set of eigenfunctions,
but different eigenvalues. The KS band structure is simply the
{\it scaled} HF one. Since
$\tau_{\rm x}$$\geq$$0$ [see Eq.(\ref{eqn:taux})], the scaling is in fact
a {\it contraction}, unless $v$$=$$0$. Although such a trivial scaling
scenario is a crude simplification of the situation realized
in realistic calculations, it does
explain the main tendency: not only the band gaps, but also the bandwidths
appear to be smaller in the KS picture. \cite{comparison}
This is definitely different from the rigid upward shift of the conduction
bands, underlying the ''scissor operator'', which affects only the
gap width.

\section{One-particle excitations}
\label{sec:excitations}

  In this section we establish several useful connections between
Kohn-Sham eigenvalues and true one-particle excitation energies
for the extended Hubbard Hamiltonian in the mean-filed approach.
First, we will considered the conductivity gap as an example of
excitations for
which the total
number of particles is not conserved.
We will show that although in this case the physical gap can be obtained
from the total energy functional
and is entirely defined by the parameters of the ground state,
the one-electron KS band structure plays
a formally fictitious role, in the sense that the KS gap
is generally different from the
true conductivity gap.
Second, we will turn to the analysis of virtual spin-wave
excitations in the antiferromagnetic insulating state
at half filling
and argue that these excitations can be
interpreted {\it directly} in terms of the KS eigenvalues.

\subsection{Conductivity gap}

  If $n_e$$=$$n_0$, the KS scheme yields the insulating solution.
The band gap $\Delta_{\rm KS}$$=$$U_{\rm KS} \tau_{\rm z}$ is direct and
opens in the points $\{{\bf k}_0\}$ for which $f({\bf k}_0)$$=$$0$
and $\vartheta^\uparrow_{{\bf k}_0}$$=$$0$ [see Eqs. (\ref{eqn:eigenv}) and
(\ref{eqn:angle})]. In order to calculate the physical conductivity gap,
we first take the standard procedure \cite{Stadele}
and employ the fact that two Hamiltonians, $\widehat{\cal H}_{\rm KS}^\uparrow$
and $\widehat{\cal H}_{\rm HF}^\uparrow$, have the same set of eigenfunctions.
Then, the conductivity gap is given by
\begin{equation}
\Delta=\Delta_{\rm KS} + \langle \psi^\uparrow_{{\bf k}_0+}|
\widehat{\cal H}_{\rm HF}^\uparrow - \widehat{\cal H}_{\rm KS}^\uparrow|
\psi^\uparrow_{{\bf k}_0+} \rangle -
\langle \psi^\uparrow_{{\bf k}_0-}|
\widehat{\cal H}_{\rm HF}^\uparrow - \widehat{\cal H}_{\rm KS}^\uparrow|
\psi^\uparrow_{{\bf k}_0-} \rangle.
\label{eqn:gap1}
\end{equation}
Taking this as an exercise and using Eq.(\ref{eqn:eigenf}) for
$\vartheta^\uparrow_{{\bf k}_0}$$=$$0$, it is very easy to verify that
$\Delta$$=$$U_{\rm HF}\tau_{\rm z}$$\equiv$$\Delta_{\rm HF}$,
which is nothing but the consequence of well-known Koopmans' theorem for
the eigenvalues of the HF Hamiltonian (\ref{eqn:HHF}).
(Note, that the correction to Koopmans' theorem due to the relaxation of
HF levels vanishes for extensive systems when $N$$\rightarrow$$\infty$).

  The central question is
how the physical gap is related with the ground-state properties,
and whether it can be
entirely described by the parameters of the ground state and the KS eigenvalues.
As it was discussed in the introduction, the most prominent scenario
is based on the idea of discontinuity of the KS potential.
In the case of finite species
the similar problem was investigated for
the chemical hardness (half of the band gap in the limit
$N$$\rightarrow$$\infty$).
It was argued that the exact KS potential
exhibits a constant jump (discontinuity) when the
number of electrons changes through an integer value, and the chemical
hardness can be found as the difference of highest occupied KS eigenvalues calculated
for the initial and the final electron configurations (see, e.g.,
Refs. \onlinecite{Perdew1,Gross,Solovyev1}).
The constant jump of the potential, however, is not related with the
ground-state properties. Similar arguments can be applied for infinite
systems, where the band gap can be expressed through the change of the
chemical
potential. \cite{GSS2}
In the context of the present work, this would be equivalent to the
claim that the uniform part
of the KS potential ($C$), which was irrelevant to the search of the
ground state and dropped out the analysis in Sections \ref{sec:OEP}-\ref{sec:HF},
should exhibit the finite jump across the band gap:
$C|_{n_0+\eta}$$-$$C|_{n_0-\eta}$$=$$\Delta_{\rm HF}$$-$$\Delta_{\rm KS}$,
and the band-gap problem might be resolved via the scissor-type correction.

\begin{figure}
\centering \noindent
\resizebox{12cm}{!}{\includegraphics{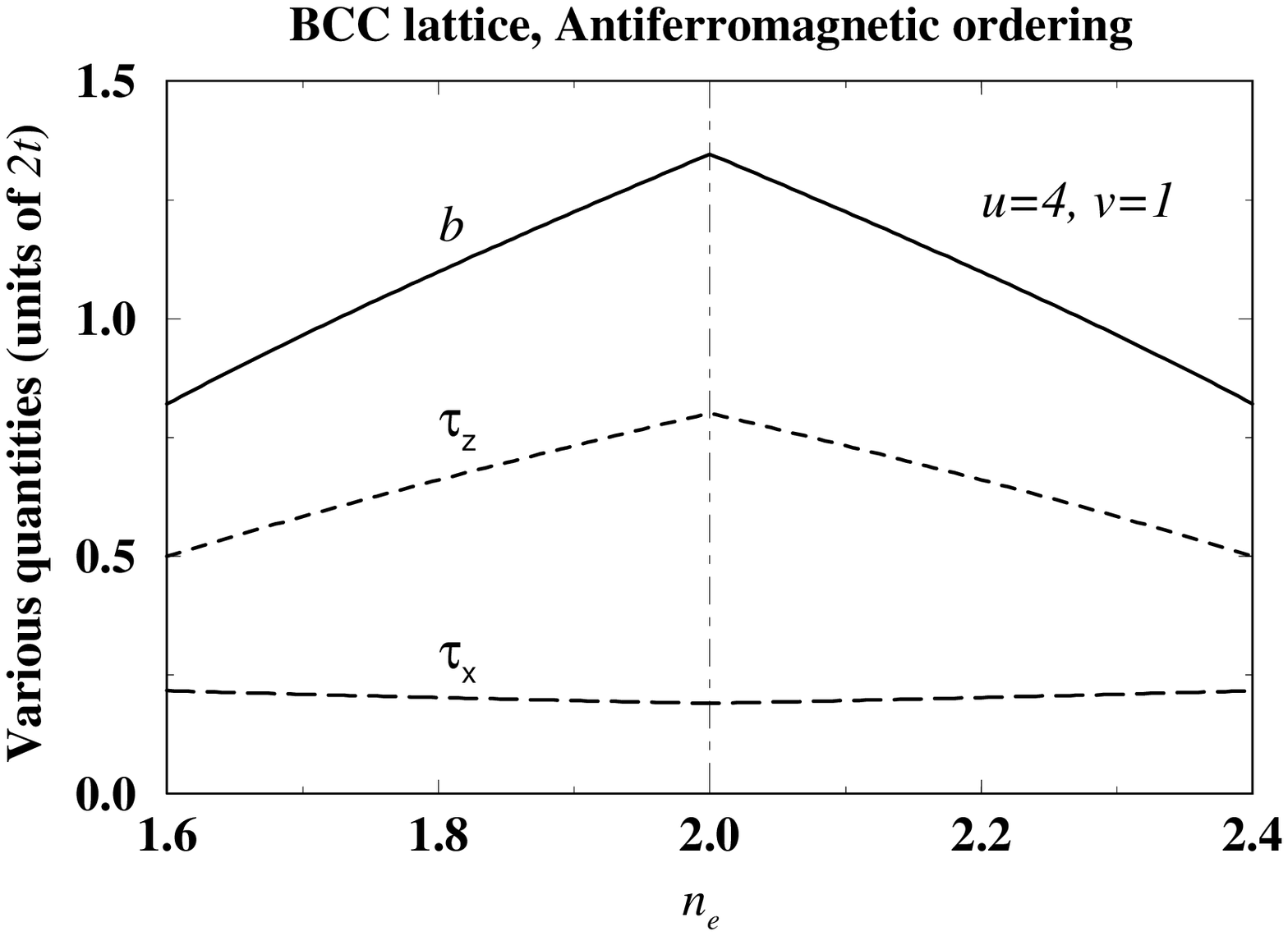}}
  \caption{Characteristic behavior of $\tau_{\rm x}$, $\tau_{\rm z}$
           and $b$ calculated self-consistently as the function of the
           electron number near $n_e$$=$$n_0$.}
  \label{fig.afbcc}
\end{figure}

  Below we present a different view on the band-gap problem.
To begin with
we note that the quantities $\tau_{\rm x}$, $\tau_{\rm z}$ and $b$,
which do determine the ground state, exhibit no finite discontinuity as the
function of the electron number (Fig.\ref{fig.afbcc}). The analysis
presented in preceding sections clearly shows that
the physical gap $\Delta_{\rm HF}$ can be expressed in terms of these
characteristics of the ground state by using another, scaling transformation
$\Delta_{\rm HF}$$=$$(1$$+$$v\tau_{\rm x})\Delta_{\rm KS}$. \cite{remark4}
Thus, {\it there is} a unique connection between $\Delta_{\rm HF}$ and the
ground-state properties.
In order to gain more insight into the problem, we argue below that
the physical gap $\Delta_{\rm HF}$ can be naturally derived from the
mean-field total energy functional ${\cal E}_{\rm MF}[\tau_{\rm x},\tau_{\rm z}]$,
not exploiting the idea of the finite discontinuity
of the exact KS potential. Instead, we argue that the band-gap problem is
related with the derivative discontinuity of the basic variable $\tau_{\rm z}$
across the band gap. We start with the general definition:
\begin{equation}
\Delta={\cal E}_{\rm MF}(n_0N+1)+{\cal E}_{\rm MF}(n_0N-1)-2{\cal E}_{\rm MF}(n_0N).
\label{eqn:gap2}
\end{equation}
As in Eq.(\ref{eqn:gap1}), we assume that the addition or the removal of an
electron does not destroy the symmetry of the ground state.
(For example, if the ground state of the $n_0N$ electron system
is antiferromagnetic,
we assume that the system of $n_0N$$\pm$$1$ electrons will remain in the
antiferromagnetic state, though this requirement may formally contradict
to the Nagaoka theorem \cite{Nagaoka}).
Then, in the limit $N$$\rightarrow$$\infty$, Eq.(\ref{eqn:gap2}) becomes
\begin{equation}
\Delta=\lim_{\eta \rightarrow 0}
\left\{ \left. \frac{dE_{\rm MF}[\tau_{\rm x},\tau_{\rm z}]}{dn_e} \right|_{n_0 + \eta}
- \left. \frac{dE_{\rm MF}[\tau_{\rm x},\tau_{\rm z}]}{dn_e} \right|_{n_0 - \eta}
\right\},
\label{eqn:gap3}
\end{equation}
where $\eta$$=$$1/N$.
Taking into account the explicit form of $\tau_{\rm x}$ and $\tau_{\rm z}$,
Eqs. (\ref{eqn:taux}) and (\ref{eqn:tauz}), we will have two contributions
to
$(dE_{\rm MF}[\tau_{\rm x},\tau_{\rm z}]/dn_e)|_{n_0 \pm \eta}$:
one is due to the change of the the band filling with the fixed
potential $b$ (the derivative of the $\Theta$-function),
and the other one is due to the $n_e$-dependence of the
self-consistent potential
itself.
Therefore, we can write:
$$
\left.
\frac{dE_{\rm MF}[\tau_{\rm x},\tau_{\rm z}]}{dn_e} \right|_{n_0 \pm \eta}
= \left.
\frac{\partial E_{\rm MF}[\tau_{\rm x},\tau_{\rm z}]}{\partial n_e} \right|_{b, n_0 \pm \eta}
+ \left.
\frac{\partial E_{\rm MF}[\tau_{\rm x},\tau_{\rm z}]}{\partial b} \right|_{n_0}
\left. \frac{\partial b}{\partial n_e} \right|_{n_0 \pm \eta}.
$$
However, in the ground state it holds
$(\partial E_{\rm MF}[\tau_{\rm x},\tau_{\rm z}]/\partial b)|_{n_0}$$=$$0$
[see Eq.(\ref{eqn:OEP})], and we find:
\begin{equation}
\Delta=\lim_{\eta \rightarrow 0}
\left\{ \left. \frac{\partial E_{\rm MF}[\tau_{\rm x},\tau_{\rm z}]}
{\partial n_e} \right|_{b,n_0 + \eta}
- \left. \frac{\partial E_{\rm MF}[\tau_{\rm x},\tau_{\rm z}]}{\partial n_e} \right|_{b,n_0 - \eta}
\right\}.
\label{eqn:gap4}
\end{equation}
Then, we have the following identity:
$(\partial\Theta[\varepsilon_F$$-$$\varepsilon^\uparrow_\pm({\bf k})]/\partial n_e)
|_{b,n_0\pm\eta}$$=$$\delta[\varepsilon_F|_{b,n_0\pm\eta}$$-$$\varepsilon^\uparrow_\pm({\bf k})]
(\partial \varepsilon_F / \partial n_e)|_{b,n_0\pm\eta}$.
In the limit $\eta$$\rightarrow$$0$, the $\delta$-function
will pick up only those points in the Brillouin zone which
correspond to the band-gap edges, i.e., ${\bf k} \in \{{\bf k}_0\}$.
Therefore, we obtain from
Eq.(\ref{eqn:efermi}):
$$
\lim_{\eta \rightarrow 0}
\left.
\sum_{\bf k} \frac{\partial\Theta[\varepsilon_F-\varepsilon^\uparrow_\pm({\bf k})]}
{\partial n_e}\right|_{b,n_0\pm\eta} \equiv
\lim_{\eta \rightarrow 0}
\left.
\sum_{{\bf k} \in \{{\bf k}_0\}}
\frac{\partial\Theta[\varepsilon_F-\varepsilon^\uparrow_\pm({\bf k})]}
{\partial n_e}\right|_{b,n_0\pm\eta} = \frac{N}{n_0}.
$$
This immediately yields:
$\lim_{\eta \rightarrow 0} (\partial \tau_{\rm x}/
\partial n_e) |_{b, n_0 \pm \eta}$$=$$0$ [see Eq.(\ref{eqn:taux})
and note that
$f({\bf k}_0)$$=$$0$], meaning that there is no
derivative discontinuity of the kinetic energy (\ref{eqn:Ekin})
across the band gap caused by the change of the band filling with the
fixed KS potential.
(Do not mix it with the behavior of $\tau_{\rm x}$ in Fig.\ref{fig.afbcc},
which shows the total effect, including the change of the KS potential
$b$).

  On the contrary, the derivative of the basic variable
$\tau_{\rm z}$ shows the discontinuity:
$n_0\lim_{\eta \rightarrow 0} (\partial \tau_{\rm z}/
\partial n_e) |_{b, n_0 \pm \eta}$$=$$\mp$$1$.
Then, putting these results  in Eq.(\ref{eqn:gap4}) and using
the explicit form of $E_{\rm MF}[\tau_{\rm x},\tau_{\rm z}]$,
Eqs.(\ref{eqn:Ekin})-(\ref{eqn:EX}), we again arrive at
Koopmans' theorem for the HF gap:
$\Delta$$=$$U_{\rm HF}\tau_{\rm z}$$\equiv$$\Delta_{\rm HF}$.

  Thus, although the KS gap is generally different from
the true conductivity gap,
$\Delta_{\rm HF}$ in our model, the latter can still be derived from
the total energy functional $E_{\rm MF}[\tau_{\rm x},\tau_{\rm z}]$
using the definition (\ref{eqn:gap4}).

\subsection{Virtual spin-wave excitations}

  Let us consider the energies of spin-wave excitations in the
antiferromagnetic insulating state at half filling.

  For these purposes
it is convenient to work with the local representation for the
one-electron potential provided by the KS scheme.
Then, the total energy changes caused by small nonuniform rotations
of the spin magnetic moments near the (antiferromagnetic) equilibrium,
i.e., the processes relevant to the spin-wave excitations,
can be exactly mapped onto the
Heisenberg-type model \cite{Liu,Liechtenstein,Solovyev2}
\begin{equation}
E[\{{\bf e}_{\it i}\}]=-\frac{1}{2}\sum_{\it ij} J_{\it ij}
{\bf e}_{\it i} \cdot {\bf e}_{\it j},
\label{eqn:Heisenberg}
\end{equation}
where  ${\bf e}_{\it i}$ is the
direction of the spin magnetic moment at the site ${\it i}$.
The parameters
of the model can be expressed
in terms of the
one-electron Green function $G^{\uparrow, \downarrow}_{\it ij}(\varepsilon)$
constructed for the KS Hamiltonian (\ref{eqn:HKS})
as \cite{Liu,Liechtenstein}
\begin{equation}
J_{\it ij}=\frac{\Delta_{\rm ex}^{\it i} \Delta_{\rm ex}^{\it j}}{2\pi} {\rm Im}
\int_{-\infty}^{\varepsilon_F} d\varepsilon
G^\uparrow_{\it ij}(\varepsilon)G^\downarrow_{\it ji}(\varepsilon),
\label{eqn:Jij}
\end{equation}
where $\Delta_{\rm ex}^{\it i}$$\equiv$$B_\nu^\downarrow$$-$$
B_\nu^\uparrow$ is the local exchange splitting at the site
${\it i}$$\equiv$$({\bf R}\nu)$.
The mapping is based on so-called local force theorem.
One can argue rather generally that if the spin-magnetization
density ($\pm$$\tau_{\rm z}$ in our model for the antiferromagnetic state)
is the basic variable in a (spin)-dinsity-functional theory, and is
requested to come out exactly of the one-electron KS orbitals,
such a mapping onto the Heisenberg-type model should be exact. \cite{Solovyev2}

  For two nonequivalent sites of our model we have:
$\Delta_{\rm ex}^{{\bf R}1}$$=$$-$$\Delta_{\rm ex}^{{\bf R}2}$$=$$U_{\rm KS}\tau_{\rm z}$.
Thus, the strength of the interatomic exchange couplings $J_{\it ij}$  is
directly related with the magnitude of the KS gap
$\Delta_{\rm KS}$$=$$U_{\rm KS}\tau_{\rm z}$.

  Let us consider the large-$u_{\rm KS}$ limit.
In the lowest order of the $1/u_{\rm KS}$
expansion, the interatomic magnetic interactions are restricted by the
nearest neighbors $\langle {\it ij} \rangle$. Corresponding elements of
the Green function are given by
$G^\uparrow_{\langle {\it ij} \rangle}(\varepsilon)$$=$$
G^\downarrow_{\langle {\it ji} \rangle}(\varepsilon)$$\simeq$$-$$
t/(\varepsilon^2$$-$$\frac{1}{4}U_{\rm KS}^2\tau_{\rm z}^2)$.
Putting them in Eq.(\ref{eqn:Jij}) and noting that in the large-$u_{\rm KS}$ limit
$\tau_{\rm z}$$=$$1$ and
$U_{\rm KS}$$=$$U$$-$$V$ [see Eq.(\ref{eqn:inflimit})], we arrive at the well-known
result for the effective magnetic coupling
between nearest neighbors in the extended Hubbard model:
\cite{Hirsch,Brink,Meinders,Eder}
$J_{\langle {\it ij} \rangle}$$=$$t^2/(U$$-$$V)$, or
$J_{\langle {\it ij} \rangle}$$=$$t^2/\Delta_{\rm KS}$.
[Note, that the sign of $J_{\langle {\it ij} \rangle}$ depends on the
convention used for the directions of the spin magnetic moments in
Eq.(\ref{eqn:Heisenberg}). If we refer to the antiferromagnetic
ground state, ${\bf e}_{\it i} \cdot {\bf e}_{\it j}$$=$$-$$1$, and
therefore $J_{\langle {\it ij} \rangle}$$>$$0$].
Thus,
the interatomic magnetic interactions can be regarded
as
virtual hoppings between effective KS levels, and in this sense the KS
band structure has the clear physical meaning.

\

  Finally, we would like to mention one rather interesting implication of two
examples discussed in this section to the theory of strongly-correlated
materials. The fact that the intersite Coulomb interaction $V$ ''renormalizes''
primarily the exchange coupling $J_{\langle {\it ij} \rangle}$ in the
antiferromagnetic state, but not the conductivity gap was well-known from
earlier studies of the extended Hubbard model. Such an apparently different
influence of the intersite Coulomb interaction $V$ on different physical
quantities has led to some ambiguity in the determination of the physical
parameter, which controls the strength of correlation effects in solids
(see, e.g., discussions in Refs.\onlinecite{Brink,Meinders}).
The present work suggests that these two properties are not necessarily
independent. In the mean-field approach, such a behavior means that
that there are (at least) two different representations for the one-particle
spectrum, which correspond to the same ground state. One representation is
essentially nonlocal, but allows the spectroscopic interpretation based on
Koopmans' theorem (the Hartree-Fock scheme). Another representation is local
and can be directly applied to the analysis of interatomic magnetic
interactions (the Kohn-Sham scheme).

\section{Summary}
\label{sec:conclusion}

  Being based on the one-band extended Hubbard model,
we have discussed several
aspects of the Kohn-Sham formulation of the density-functional
theory in solids.
Although the conclusions are strictly applicable only to the mean-filed
solutions of this particular model, the analysis presented in the paper
suggests that the old-standing dilemma between the description of the
ground-state properties in the density-functional theory and the description
of (at least some) one-particle excitations may be resolved in a more rational
way. Particularly, we have shown that the conductivity gap and the interatomic
magnetic interactions relevant to the spin-wave excitations
for this model are entirely defined by
the parameters of the ground state. This therefore means that in this case
there is no fundamental difference between the ground-state properties,
the conductivity gap and the spin-wave excitation energies,
and all of them can be described by the
same set of one-electron Kohn-Sham equations.
The KS eigenvalues may not be necessarily
coincide with the true one-particle excitation energies,
depending on the form in which the nonlocal exchange interactions contribute
to each particular type of excitations.

  Thus, we hope that results of our model analysis may
provide a new insight into the interpretation of one-electron equations underlying
the Kohn-Sham scheme and stimulate new researches in this direction.
As an extension of the present analysis, we believe that it would be
particularly interesting:
(i) to go beyond the mean-field approximation and to take into account
the correlation effects;
(ii) to go beyond the one-band picture, to take into account the effects
of orbital
degeneracy and to
clarify
the roles played by intersite Coulomb interactions in the problem of
orbital polarization
in solids.\cite{Solovyev3}

\section*{Acknowledgments}

I thank K. Terakura and F. Aryasetiawan for useful discussions.
The present work is partly supported
by New Energy and Industrial Technology
Development Organization (NEDO).

\begin{table}
\caption{Symmetry constraints superimposed on the
density matrix and the KS potential.
The uniform potential shift is set as zero.
Note also that the number of
electrons is $n_e$$=$$n_0{\rm Tr}(\widehat{\rho}^\uparrow)$.
In the COF state, the minority-spin bands are requested to be
unoccupied; in all other respects the parameters
$B^\downarrow_{1,2}$ can be arbitrary.}
\label{constraints}
\begin{center}
\begin{tabular}{ccc}
state & density matrix    & KS potential        \\
\tableline
AF    & $\widehat{\rho}^\downarrow$$=$$\widehat{\sigma}_{\rm x}\widehat{\rho}^\uparrow\widehat{\sigma}_{\rm x}$ &
$-$$B_1^\uparrow$$=$$B_1^\downarrow$$=$$B_2^\uparrow$$=$$-$$B_2^\downarrow$ \\
CON   & $\widehat{\rho}^\downarrow$$=$$\widehat{\rho}^\uparrow$ &
$-$$B_1^\uparrow$$=$$-$$B_1^\downarrow$$=$$B_2^\uparrow$$=$$B_2^\downarrow$ \\
COF   & $\widehat{\rho}^\downarrow$$=$$0$ &
$-$$B_1^\uparrow$$=$$B_2^\uparrow$ \\
\end{tabular}
\end{center}
\end{table}


\begin{thebibliography}{10}
\bibitem{remark1}
Note, that the antiferromagnetism or the charge ordering implies
the existence of
magnetic or charge inhomogeneities
on the interatomic scale. Most part of the first-principles
DFT methods is based either on the purely homogeneous electron gas
picture for the exchange-correlation energy (LSDA), or takes into
account only local effects of the inhomogeneity through gradient
expansions \cite{KohnSham,ParrYang}.
\bibitem{remark3}
Although we do not consider here the situation, when the insulating
ground state is caused by chemical bonding effects, the extension
of our arguments to this case is rather straightforward.
\bibitem{remark4}
The scaling transformation
naturally explains that $\Delta_{\rm HF}$$\ge$$\Delta_{\rm KS}$ when
$v$$\ge$$0$. In fact, a similar tendency for the KS gap and the true
conductivity gap $\Delta$ was obtained by Sch\"{o}nhammer and Gunnarsson.
\cite{SchGun} They considered a formally exact KS scheme for a
Hubbard-type model, which was solved numerically for finite systems with
subsequent extrapolation of results to $N$$\rightarrow$$\infty$.
They found that $\Delta$$\approx$$\Delta_{\rm KS}$ only when
$v$$=$$0$, whereas more generally it held $\Delta$$\ge$$\Delta_{\rm KS}$.
They interpreted the difference between $\Delta$ and $\Delta_{\rm KS}$
using the discontinuity scenario (see also Ref.\onlinecite{comment}).
Our standpoint is different.
\end{thebibliography}
\end{document}